\documentclass[preprint,showpacs,amsmath,amssymb,prl,superscriptaddress,
floatfix]{revtex4}

\usepackage{graphicx}
\usepackage{dcolumn}
\usepackage{bm}

\begin{document}

\title{Scattering Dynamics of Driven Closed Billiards}

\date{\today}


\author{Florian Lenz}

\email[]{lenz@physi.uni-heidelberg.de}

\affiliation{Physikalisches Institut, Universit\"at Heidelberg, Philosophenweg 12, 69120 Heidelberg, Germany}%
\author{Fotis K. Diakonos}
\affiliation{Department of Physics, University of Athens, GR-15771 Athens, Greece}
\author{Peter Schmelcher}


\affiliation{Physikalisches Institut, Universit\"at Heidelberg, Philosophenweg 12, 69120 Heidelberg, Germany}%

\affiliation{Theoretische Chemie, Institut f\"ur Physikalische Chemie, Universit\"at Heidelberg, INF 229, 69120 Heidelberg, Germany}%

\date{\today}


\begin{abstract}\label{ch:abstract}
We investigate the classical scattering dynamics of the driven elliptical billiard.
Two fundamental scattering mechanisms are identified and employed to understand the
rich behavior of the escape rate. A long-time algebraic decay which can be tuned
by varying the driving amplitude is established. Pulsed escape rates and
decelerated escaping  particles are generic properties of the harmonically
breathing billiard. This suggests time-dependent billiards as prototype systems
to study the nonequilibrium evolution of classical ensembles encountering a
multitude of scattering processes off driven targets.
\end{abstract}

\maketitle

Many models of statistical mechanics can be reduced to
billiard-type dynamical
systems. Seminal results like the justification of a probabilistic
approach to statistical mechanics ~\cite{Zas99, Ego00} or a connection between
the escape from a billiard and the famous Riemann hypothesis ~\cite{Bun05} were
obtained.
Modern studies of billiards are, among others, motivated by corresponding
experiments including ultracold atoms confined in a laser potential 
~\cite{Rai01, Dav011, And04, And06}, microwave billiards ~\cite{Sto99, Gra92,
Ste92} or mesoscopic
quantum dots ~\cite{Mar92}. Equally for the design of directional
micro-lasers, billiards are of immediate relevance ~\cite{Noc97}. From the
theoretical
point of view, investigations on the classical and quantum properties of
billiards have pioneered the fields of quantum chaos, modern semiclassics and
transport at the mesoscopic scale (see ~\cite{Sto99} and refs. therein). 

The escape rate, being the fraction of remaining particles as
a function of time, is a characteristic of open billiards which is both
experimentally accessible and important
for transport properties ~\cite{Bun05}. This key property allows
to probe
the dynamics from the outside and has been studied thoroughly for billiards with
a static boundary ~\cite{Bau90, Alt96, Kok00}. Integrable  systems exhibit an
algebraic, while fully chaotic billiards show an exponential decay of
the escape rate ~\cite{Bau90} (altough it is known that in special cases,
like the Bunimovich stadium, there is an algebraic long-time behavior in the
decay, due to the slow transport of particles close to the marginally stable
bouncing ball orbits, cf. e.g. ~\cite{Alt96}). Here we focus
on
the case where the billiard shape changes in time according to a certain law.
Such a driven billiard not only leads to a higher dimensional phase space of the
scattering processes but also to a non-conservation i.e. time-evolution of the
energy.
Driven billiards represent prototype systems for the evolution of ensembles of
particles
in a closed driven environment where multiple scattering off the driven boundary
takes place 
thereby leading to a dynamical non-equilibrium state.
Extremely little is known on the properties of such systems, the 
few existing investigations dealing with aspects such as Fermi acceleration
~\cite{LL92, Los99, Los02, Sch06}
and principal structures of the corresponding phase space ~\cite{Koi95b, Koi95,
Iti03}.
In spite of the above-motivated interest escape rates of driven billiards have
not been addressed up to date.
Using atom-optical techniques such as acousto-optic deflectors or even in the
case of microwave cavities moving boundaries are well within reach of
experiment.

We investigate the classical scattering dynamics and the time-evolution of
ensembles
of particles in a harmonically driven elliptical billiard.
The decay of the escape rate is traced back to the underlying scattering
mechanism by identifying two fundamental scattering processes being key
ingredients
for the time-evolution. The escape rate behaves asymptotically as $N_C(t) \sim
t^{-w_C}$ 
and it is shown that the decay constant $w_C$ can be changed
continuously by varying the driving amplitude $C$.
Moreover, phenomena such as a pulsed escape, demonstrate the richness of the
properties
of driven billiards and suggest specifically the driven ellipse as a prototype
system
for the nonequilibrium evolution of ensembles that experience multiple
scattering processes
with moving targets.

As a precursor
for the driven system let us discuss some relevant features of the static
elliptical
billiard. The boundary $\mathcal{B}$ of an ellipse is given by
$
   \mathcal{B} = \left\{ (x(\varphi) = A  \cos \varphi,y(\varphi) = B  \sin
\varphi)| 0 \le \varphi <
    2\pi  \right\}
$
with $A>B>0$, thus $A$ and $B$ being the long and the
short half-diameter, respectively.  The dynamics in the ellipse is completely
integrable ~\cite{Koi95b}, see the Poincar\'e surface of section (PSS)
shown in Fig. ~\ref{fig:ElPSS}. In addition to the energy, there is a
second constant of motion
\begin{equation}\label{eq:El2ndIntPhi}
    F(\varphi,p)= \frac{p^2 (1+ (1-\varepsilon^2) \cot^2
\varphi)-\varepsilon^2}{1+ (1-\varepsilon^2) \cot^2 \varphi -
    \varepsilon^2},
\end{equation}
restricting the orbits to invariant curves in phase space, where $\varepsilon =
\sqrt{1-{B^2}/{A^2}}$ is the eccentricity and $p=\cos \alpha$
($\alpha$ is the angle between the tangent at the boundary and the trajectory of
the particle).
$F(\varphi,p)$ can be
interpreted as the product of the angular momenta (PAM) about the two foci
~\cite{Ber81}. There are two different types of orbits in the
ellipse, librators and rotators, see Fig. ~\ref{fig:ElPSS}. Librators cross the
x-axis between the two foci and touch repeatedly a confocal hyperbola.
They possess values of $F$ between $-\varepsilon^2/(1-\varepsilon^2)$ and zero
and
their motion is restricted to a limited range of $\varphi$. Rotators travel
around the ellipse, eventually exploring every value of $\varphi$,  repeatedly
touching a confocal ellipse. They have values of $F$ between zero and one.

\section{Escape rate} 
\begin{figure}[ht]
\includegraphics[width=0.9\columnwidth]{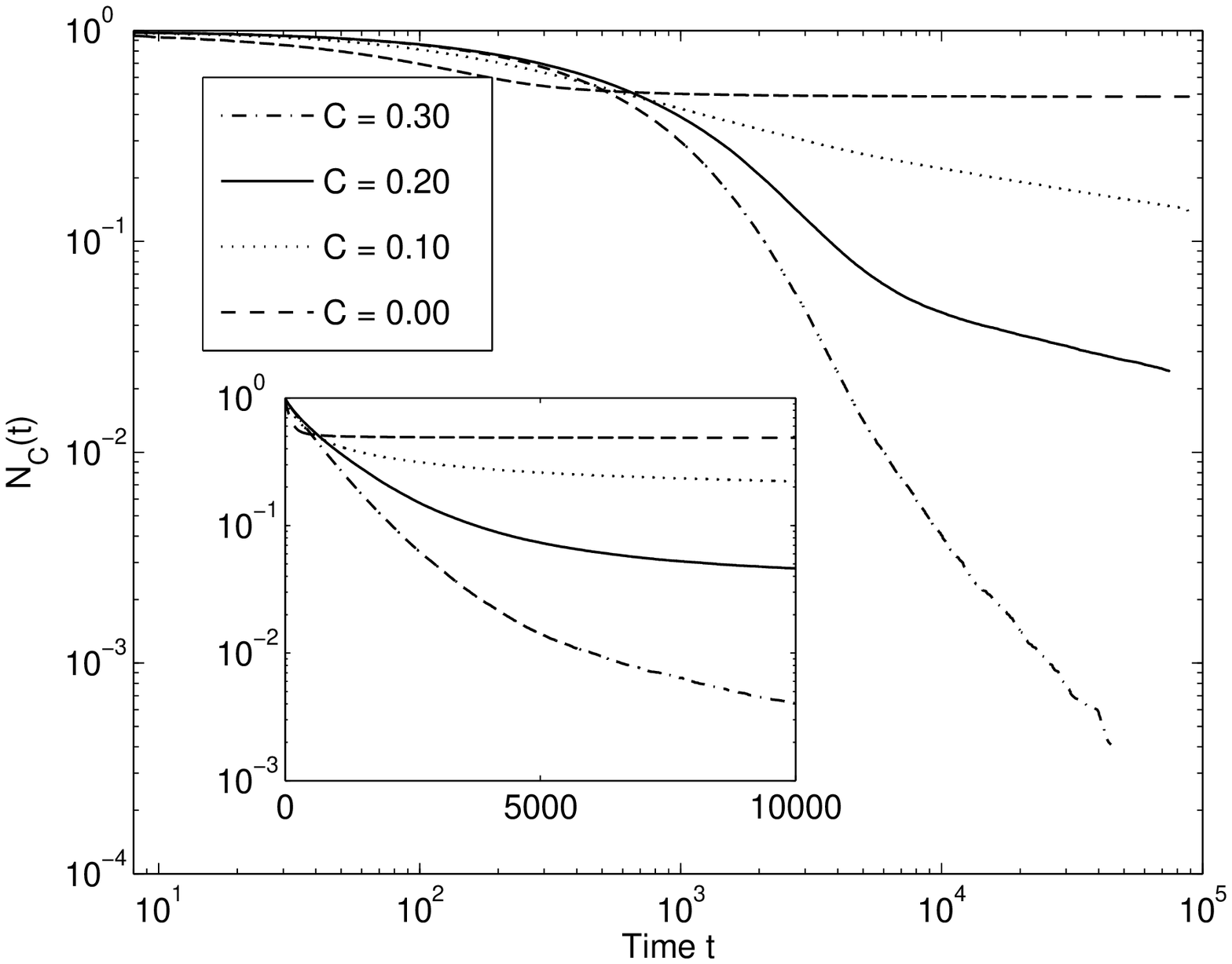}
        \caption{Fraction of remaining particles $N_C(t)$ of the IVE as a
function of time for different values of the driving amplitude $C$. The
inset provides a semi-logarithmic plot.}\label{fig:EscSL_IV}
\end{figure}

Let us consider the escape rate of the static billiard by
placing a small hole
at the very right ($\varphi =0$) of the ellipse ($A=2,\,B=1$ in arbritrary
units) and iterating the corresponding 2D discrete
mapping (for the phase space variables $\varphi$ and $p$) numerically. We employ
an ensemble of $10^6$ particles with initial conditions being uniform randomly
distributed 
in $\varphi, \alpha$-space. The result is shown in Fig. ~\ref{fig:EscSL_IV}
(curve with $C= 0.00$). The decay
approaches a saturation value $N_s(\varepsilon)$ which is caused by particles
traveling on librator orbits that are not connected with the hole
~\cite{Dav011}. 
For $\varepsilon =0$ the saturation value is zero and it increases monotonically
with increasing $\varepsilon$. The analytical dependence of $N_s$ on
$\varepsilon$ can be obtained by 
determining the number of librator type particles starting from the region
bounded by the separatrix ~\cite{Len07} 
\begin{equation}\label{eq:ElEscSaturation}
    N_s(\varepsilon) = \frac{1}{\pi^2} \int_{0}^{2\pi} d \varphi
    \, \arccos
\sqrt{\frac{\varepsilon^2}{1+(1-\varepsilon^2)\cot^2 \varphi}}
\end{equation}
This expression is in excellent agreement with the results of the numerical
simulations.
One can therefore conclude that varying $\varepsilon$ allows us to
control the total number of
particles being emitted.

\begin{figure}[ht]
\includegraphics[width=0.8\columnwidth]{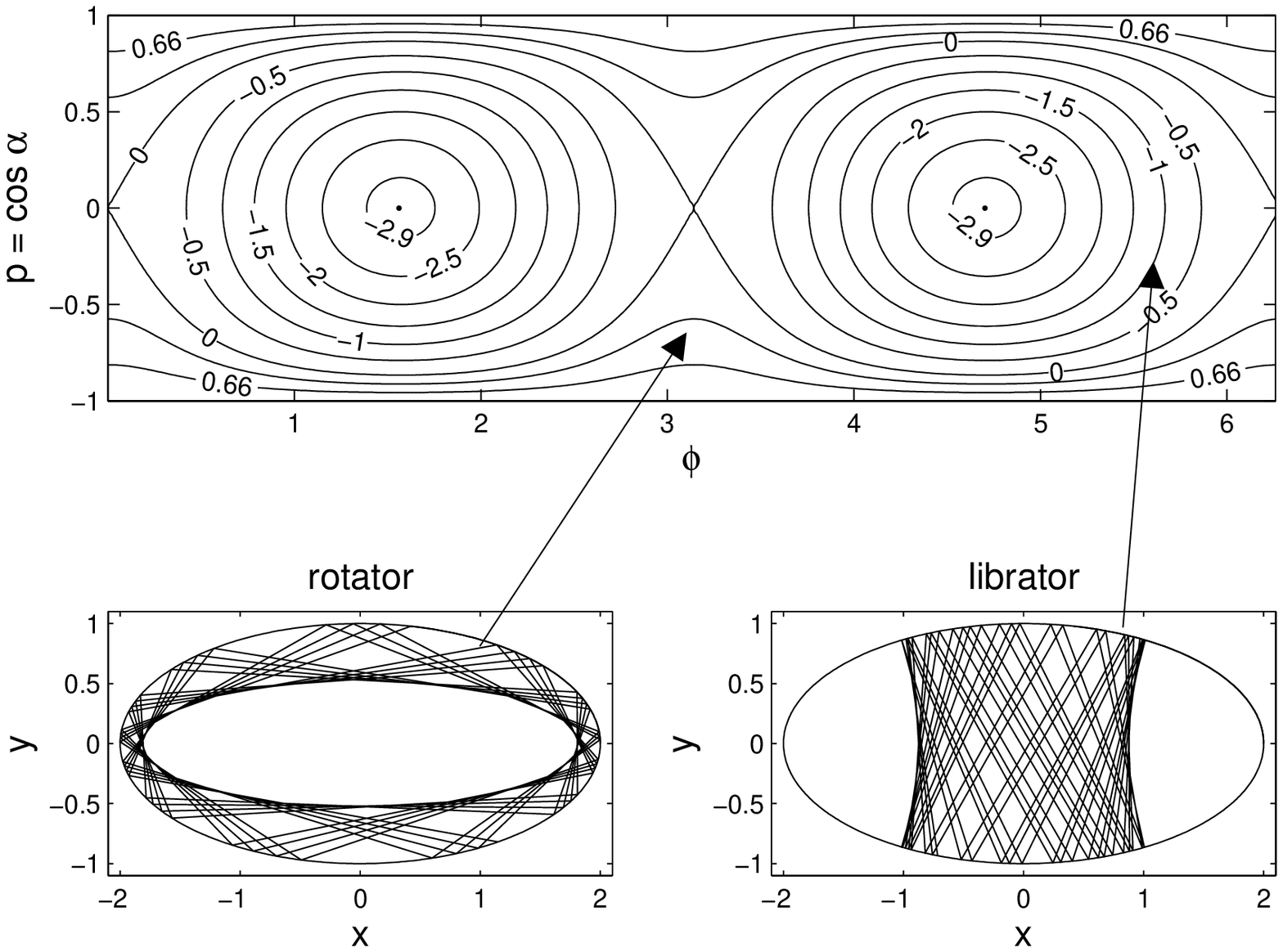}
\caption{Phase space of the ellipse, the invariant curves are the contour lines
of $F(\varphi, p)$ (upper part) and typical trajectories in coordinate space (lower part).}\label{fig:ElPSS}
\end{figure}

We now apply harmonic oscillations to the
boundary of the ellipse
\begin{equation}\label{eq:EllBoundary}
 \bm{b}(\varphi,t)=  \left (\begin{array}{c} (A_0 + C \sin (\omega t+ \delta))
\cos \varphi \\
   (B_0 + C \sin (\omega t+ \delta)) \sin \varphi\\ \end{array}\right )
\end{equation}
$C>0$ is the driving amplitude and $\delta$ is a phase shift which will be set
to zero in the following. The dynamics of the particles is now governed by a
4D discrete mapping ~\cite{Koi95b, Koi95}: the time and position of a
particle-boundary
collision is described by the pair $\varphi,t$, the energy and direction of
motion by
$\bm{v}=(v_x,v_y)$. The time $t_{n+1}$ for the $n+1$th collision is determined
implicitly by
\begin{equation}\label{eq:Implicit2}
    \left ( \frac{v_n^x (t_{n+1}-t_n) + x_n}{A_0 + C \sin (\omega t+ \delta)}
\right )^2 + \left (\frac{v_n^y (t_{n+1}-t_n) + y_n}{B_0 + C \sin (\omega t+
\delta)} \right )^2 -1 = 0,
\end{equation}
where the smallest $t_{n+1}>t_n$ that solves ~(\ref{eq:Implicit2}) has to be
taken and $\bm{x}_n=(x_n,y_n)$ is the $n$th collision point.
The $n+1$ collision point is given by $\bm{x}_{n+1}=\bm{x}_n+\bm{v}_n
(t_{n+1}-t_n)$ and $\varphi_{n+1}$ can be obtained by inverting
~(\ref{eq:EllBoundary}). The velocity immediately after the collision
$\bm{v}_{n+1}$ is
described by $ \bm{v}_{n+1} = \bm{v}_n -2 \left [ \hat{ \bm{n}}_{n+1} \cdot
(\bm{v}_n -
\bm{u}_{n+1})\right ] \cdot \hat{ \bm{n}}_{n+1}
$, where $\bm{u}_{n+1}$ is the boundary velocity and $\hat{ \bm{n}}_{n+1}$ the
inward
pointing normal vector of the collisional event occurring at time $t_{n+1}$ and
position $\bm{x}_{n+1}$. When iterating the underlying implicit
mapping numerically the determination of $t_{n+1}$ from eq.
~(\ref{eq:Implicit2}) 
requires sophisticated numerical techniques due to many neighboring roots. As a
result
the corresponding simulations are computationally very demanding.

\begin{figure}[ht]

\includegraphics[width=0.9\columnwidth]{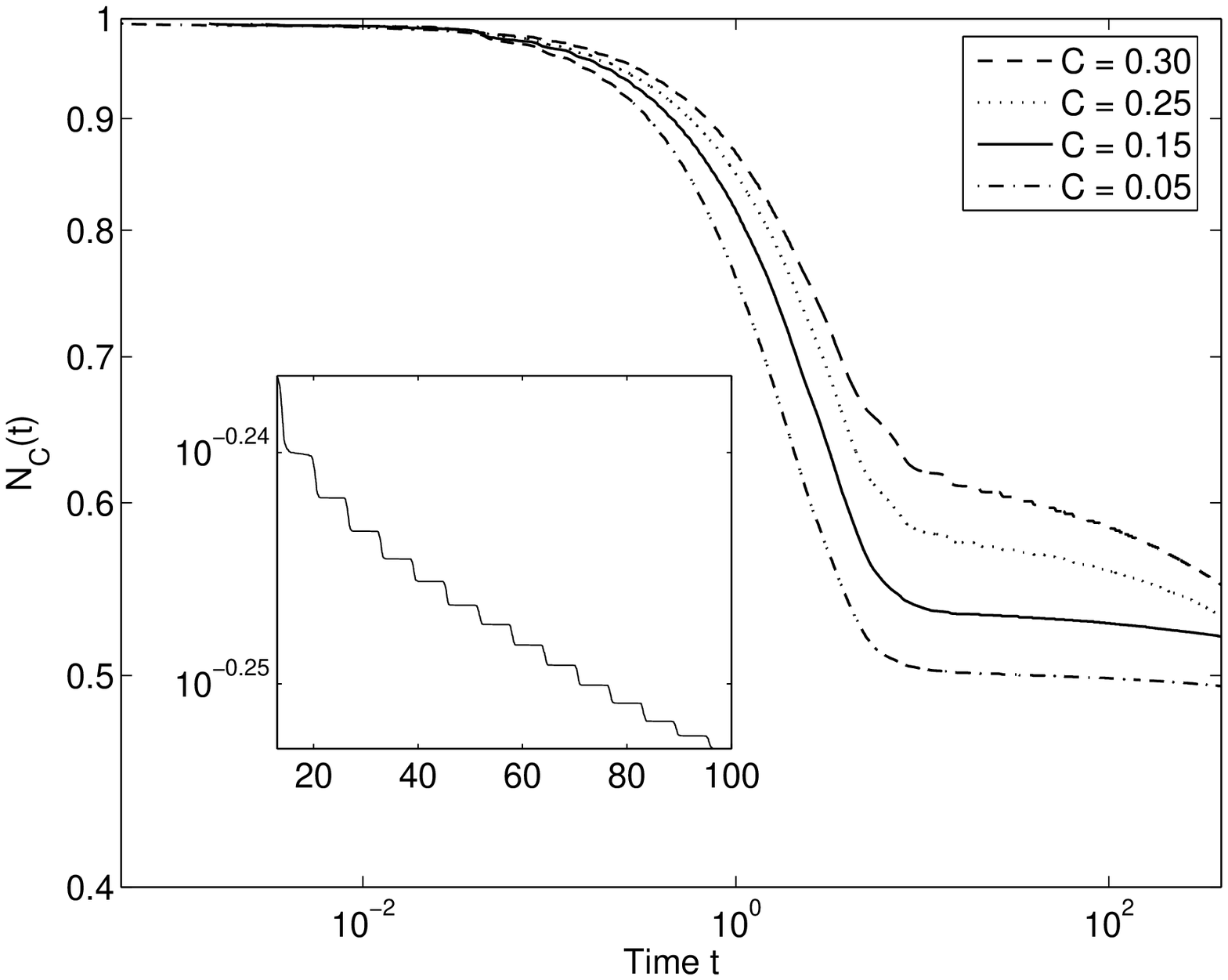}
        \caption{Fraction of remaining particles $N_C(t)$ for the HVE, 
$2\pi$-oscillations of the decay are shown in the inset.}\label{fig:EscSL_HV}
\end{figure}

We focus on the escape rates of monoenergetic ensembles
consisting of $10^5$ particles \footnote{Particles start
from the innermost ellipse, $\alpha$ is distributed uniform randomly
and $\bm{v}_0= v_0 \cdot  (\cos \alpha, \sin \alpha)$, where $v_0=1$
(IVE) and $v_0=100$ (HVE).} with $\omega =1$, $A_0=2$ and $B_0 =1$ for different
values of the
driving amplitude $C$. Two relevant cases can be distinguished:
$|\bm{v_0}|\approx
\omega C$ being the intermediate velocity ensemble (IVE) for which the initial
particle velocity
is of the order of the maximum velocity of the billiard boundary and
$|\bm{v_0}|\gg \omega C$ being the high velocity ensemble (HVE). Naturally, it
would be also interesting
to examine the case $|\bm{v}_0|\ll \omega C$. However, the first
few collisions then accelerate the particles to velocities
$|\bm{v}|\approx \omega C$ and, after a short time, we encounter
the situation of the IVE.
$N_C(t)$ of the IVE and HVE are shown in
Figs. ~\ref{fig:EscSL_IV} and ~\ref{fig:EscSL_HV}, respectively. In both
cases, there exists no saturation value as observed in the static case. A
short-time 
exponential decay is followed by a transient and the long-time behavior
($t>10^4$) is to a good approximation algebraic, $N_C(t)\sim t^{-w_C}$, at least
for the case of the IVE (we remark that this algebraic decay has been
numerically shown to exist for much longer times than illustrated in
Fig. ~\ref{fig:EscSL_IV}). The decay constant $w_C$ increases monotonically with
increasing $C$ (this fact is based not only on the four values of the drivig
amplitude $C$ shown here, but on simulations carried out for 20 values of $C$
between $0.01$ and $0.30$).
It can be therefore concluded that tuning the driving amplitude allows us
to control
the decay constant $w_C$ in the long-time behavior of the decay.
The division of the behavior of the escape rate into the above-mentioned
regimes 
is even more pronounced in the case of the HVE,
see Fig. ~\ref{fig:EscSL_HV}, although the long time tail shows substantial
deviations from an algebraic behavior.

The saturation
value $N_s(\varepsilon)$ observed in the static case 
is caused by librating orbits which are gradually destroyed by the driving
~(\ref{eq:EllBoundary}) 
thereby resulting in a non-vanishing rate $\dot{N}_C (t)$ even
for large times. Rather than trying to examine the 4D phase space (since 2D
intersections or projections  of this 4D space are nonrepresentative
and difficult to interpret), let us analyze
the destruction of the librator orbits as follows.
Orbits $(\varphi_i',p_i')$ in the driven ellipse can be compared to the
corresponding ones of the static ellipse $(\varphi_i,p_i)$ by
inspecting the angular momentum $F(\varphi,p)$. In contrast to the case of the
static
ellipse where $F(\varphi_i,p_i)=const. \,\forall \,i$, 
$F(\varphi_i',p_i') \neq F(\varphi_j',p_j') \, (i\neq j)$ in the
driven case, i.e. $F$ is no longer a constant of motion. The
difference $\triangle F=F_{i+1}-F_i$ (see Fig. ~\ref{fig:ElPSS}) upon a
collision is a
measure of whether a librator orbit approaches ($\triangle F > 0$) the
separatrix  or whether it
moves in phase space towards ($\triangle F < 0$) the corresponding elliptic
fixed points of the static system. Two fundamental scattering processes
can be identified that destroy librators by changing them into rotator orbits:
\begin{enumerate}
    \item \textit{Vertical process:} The angle of incidence of a
particle upon collision does not coincide with the reflection angle due to
momentum
momentum transfer by the moving billiard boundary. This momentum change
$\triangle p$ causes a vertical displacement of the particle in the PSS.
    \item \textit{Horizontal process:} A particle that would hit the boundary at
a certain point $\varphi$ in case of a static billiard, hits the boundary in the
driven
case at $\varphi'$, simply because it has moved. Here no change of the momentum
takes place. This corresponds to a horizontal displacement $\triangle \varphi$
of the particle
in the PSS.
\end{enumerate}
These processes are fundamental in the sense that every change $\triangle F$ can
be decomposed into $\triangle F = \triangle F_h+\triangle F_v$, where $\triangle
F_{h,v}$ denote the individual changes caused by the horizontal, the vertical
process, respectively. 

In general, these effects do not appear isolated but a combination
$(\triangle \varphi,\, \triangle p)$ of both will occur at a
single collision. The value $\triangle
F_{h,v}$ at a single collision can be calculated exactly ~\cite{Len07}. The
changes $\triangle F_{h,v}$ increase monotonically with the driving amplitude
$C$. Additionally, $\triangle F_v$ depends monotonically on the driving
frequency $\omega$. It is important to note that collisions occurring during
the expansion/contraction period of the ellipse always lead to $\triangle
F_v>0$/$\triangle F_v<0$, whereas such a clear
distinction is not possible for $\triangle F_h$.

For a single particle undergoing many scattering processes, the
effective change $\triangle F_e$ after a certain time depends on the
sequence of these processes, hence $\triangle F_e = \triangle F_1 +
\triangle F_2+ \dots +\triangle F_n$ after $n$ collisions. When regarding an
ensemble of particles, the effective changes $(\triangle F_e)_j$ (where the
index $j$ indicates the $j$th particle) after $n$ collisions can vary
significantly from particle to particle, since the sequence of the individual
$\triangle F_i$ will vary substantially for different initial conditions which
is
due to the nonlinear dynamics of the underlying discrete mapping.

\begin{figure}[ht]

\includegraphics[width=0.9\columnwidth]{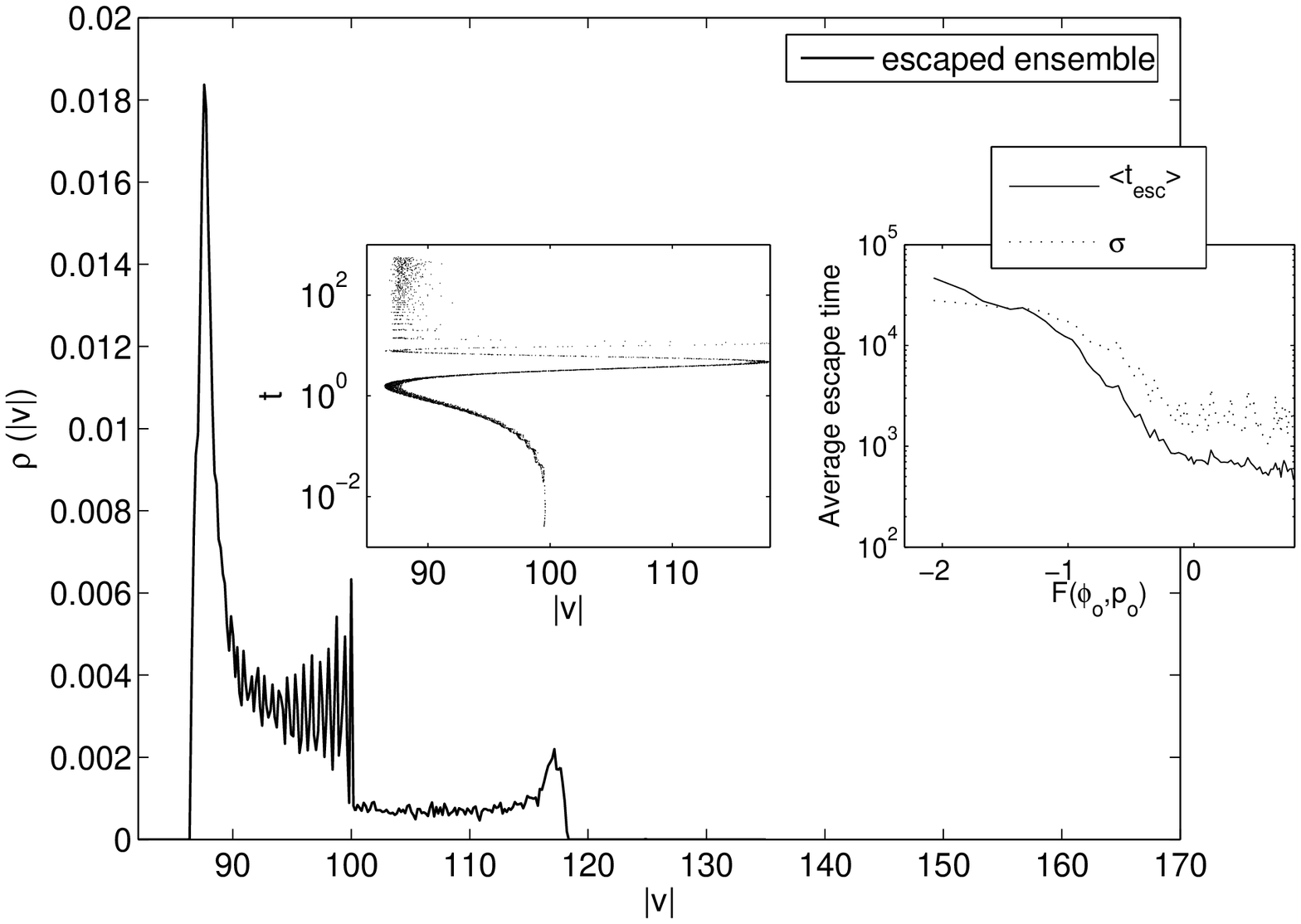}
        \caption{Distribution $\rho(|\bm{v}|)$ of the velocity of the escaped 
ensemble (HVE, $C=0.25$). Left inset: the escape time vs. the
escape velocity is plotted. Right inset: average escape time of the
IVE ($C=0.10$) as a function of the initial
$F(\varphi_0,p_0)$.}\label{fig:EsdcSL_dV}
\end{figure}

With the just presented considerations,
we can explain qualitatively the observed escape
rates and especially the disappearance of the saturation value.
Let us focus first on the HVE. The initial fast decay
($t<5$) is due to the rotator orbits that are connected with the hole (in the
PSS) and escape very rapidly.  The long-time decay ($t>10$) is
caused by particles starting on librator orbits that have been scattered
through horizontal and vertical processes across
the separatrix.  The closer a particles initial orbit lies near the elliptic
fixed points (of the static billiard), the longer it takes until the
effective change $\triangle F$ is large enough to reach the separatrix ($F=0$),
see Fig. ~\ref{fig:EsdcSL_dV}. 
The monotonic dependence of the $\triangle F_i$ on $C$  explains
the increasing emission rate $\dot{N}_C(t)$ with increasing $C$, since at a
given time $t$, the number of particles that can participate in the decay is
larger for larger values of $C$. The decay in  the transient region ($5<t<10$)
is caused by a superposition of the tail of the initial fast exponential decay 
and the onset of the slow algebraic decay. With the
same arguments, the decay of the IVE can be explained
qualitatively. Since $\mathcal{O}(\bm{v}_0) \approx \mathcal{O}(max(\bm{u}))$,
the resulting changes $\triangle F$ are much larger for the case of the IVE
compared
to the ones of the HVE. This leads to a very early onset of the slow algebraic
decay and consequently the transient region is broadened.

The above presented considerations are confirmed by the results shown in
the right inset of Fig.  ~\ref{fig:EsdcSL_dV}, where exemplarily the average
escape time and the standard deviation $\sigma$ of the IVE ($C=0.10$) as a
function of the initial PAM $F(\varphi_0,p_0)$  are plotted. Particles
with values of $F(\varphi_0,p_0)$  between zero and one (rotators) have
approximately
the same average escape times $<t_{esc}>$, but the large (compared to
$<t_{esc}>$) standard deviation $\sigma$  indicates that
the individual escape times can vary significantly from particle to particle.
The average escape time of particles starting on librator orbits
($F(\varphi_0,p_0)<0$) increases with decreasing $F(\varphi_0,p_0)$, since
the required $\triangle F_e$ to change a librator into a rotator becomes
larger and more and more collisions are necessary to reach these values of
$\triangle
F_e$. 

In the inset of Fig. ~\ref{fig:EscSL_HV}, a
modulation of the
escape rate with period $T=2\pi$ can be seen, being exactly the period
of the applied driving law ~(\ref{eq:EllBoundary}). Specifically, for
$t \ge 10$, where all particles starting on rotator orbits have already escaped,
$N_C(t) \approx const. $ during approximately $11/12$ (empirically observed) of
one period and subsequently $\dot{N}_C(t) \neq 0$
during a time interval $T/12$ only. From this behavior it is evident
that the ellipse operates from a certain time on as a pulsed source of
particles. These repeated intervals are centered around points $t_m$ of maximal
extension of the ellipse, $t_m= (4m +1)\pi/2,\, m= 2,3,4,\dots$ During the
expansion period, dominantly vertical but also horizontal processes turn
librators
into rotators. The moving ellipse remains for a comparatively long time period
in the vicinity of the extremal configuration at $t_m$ and consequently the
newly created rotators escape. Therefore, the dynamics is effectively probed
during these short time intervals centered around $t_m$.
During the contraction period, the librators are stabilized via  vertical
processes, consequently $\dot{N}_C(t)
\approx 0$ during $11/12$ of a period $T$. 

An astonishing feature of the escape of the HVE is
displayed in Fig.~\ref{fig:EsdcSL_dV},
where the distribution $\rho (|\bm{v}|)$ of the
escape velocities and the correlation of the escape time vs. the escape
velocity (left inset)
after a certain time is shown for $C=0.25$. There is a strong asymmetry of $\rho
(|\bm{v}|)$
around the initial velocity $|\bm{v}_0|=100$. Particularly, particles with
$t_{esc}>10$,
being originally exclusively librators, possess escape velocities $|\bm{v}| <
|\bm{v}_0|$. 
The origin of this behavior are vertical processes which turn librators into
rotators that subsequently escape. The latter process however takes place during
the
expanding phase of the moving ellipse which leads exclusively to a decrease i.e.
loss
of energy ~\cite{Pa02,Pa04}. The escape velocities of particles starting on
rotator orbits lie
on a serpentine band, whose oscillations match the periodic oscillations of the
ellipse.
We therefore conclude that the breathing ellipse acts as a decelerator for
particles in
the HVE regime.

In conclusion, we have explored the dynamics of a driven
elliptical billiard with
a focus on the scattering mechanisms and their impact on the escape rate.
Observed phenomena such as the long-time algebraic decay which can be tuned
by varying the driving amplitude, a pulsed escape rate and lowered velocities
of the escaping particles are all caused by projectiles emanating from librator
orbits.
They can be understood by means of two fundamental scattering processes that
turn
librating into rotating orbits. Consequently, these effects can be further
enhanced 
and manipulated by preparing suitable initial ensembles. For example, a high 
velocity ensemble initially consisting exclusively of librators restrains
particles with an escape time $t<10$ which do not show a pulsed decay and whose
escape
velocities are symmetrically distributed around the initial velocity.
It is foreseeable that the application of different driving laws and driving
modes (e.g. area-preserving
oscillations) as well as the preparation of suitable initial (thermal) ensembles
will further
advance equally our understanding as well as the phenomenology of driven
nonequilibrium
systems. Driven billiards serve as prototype systems in this respect.

Valuable discussions with A. Richter, V. Constantoudis, A. Karlis and M.
Oberthaler are gratefully acknowledged.

{}
\end{document}